\documentclass[%preprint,
superscriptaddress,
 amsmath,amssymb,
 aps,
reprint
]{revtex4-1}

\usepackage{color}
\usepackage{graphicx}% Include figure files
\usepackage{dcolumn}% Align table columns on decimal point
\usepackage{bm}% bold math
\usepackage{epstopdf}
\usepackage{gensymb}
\usepackage{float}

\begin{document}

%\preprint{APS/123-QED}

\title{Control of polymorphism during epitaxial growth of hyperferroelectric candidate LiZnSb on GaSb (111)B}% Force line breaks with \\

\author{Dongxue Du}
\thanks{These authors contributed equally.}
\affiliation{Materials Science and Engineering, University of Wisconsin-Madison, Madison, WI 53706}

\author{Patrick J. Strohbeen}
\thanks{These authors contributed equally.}
\affiliation{Materials Science and Engineering, University of Wisconsin-Madison, Madison, WI 53706}

\author{Hanjong Paik}
\affiliation{Platform for the Accelerated Realization, Analysis, \& Discovery of Interface Materials (PARADIM), Cornell University, Ithaca, NY 14853}

\author{Chenyu Zhang}
\affiliation{Materials Science and Engineering, University of Wisconsin-Madison, Madison, WI 53706}

\author{Konrad Genser}
\affiliation{Department of Physics, Rutgers University, New Brunswick, NJ 08901}

\author{Karin M. Rabe}
\affiliation{Department of Physics, Rutgers University, New Brunswick, NJ 08901}

\author{Paul M. Voyles}
\affiliation{Materials Science and Engineering, University of Wisconsin-Madison, Madison, WI 53706}

\author{Darrell G. Schlom}
\affiliation{Platform for the Accelerated Realization, Analysis, \& Discovery of Interface Materials (PARADIM), Cornell University, Ithaca, NY 14853}
\affiliation{Department of Materials Science and Engineering, Cornell University, Ithaca, NY 14853}
\affiliation{Kavli Institute at Cornell for Nanoscale Science, Cornell University, Ithaca, NY 14853}

\author{Jason K. Kawasaki}
\email{jkawasaki@wisc.edu}
\affiliation{Materials Science and Engineering, University of Wisconsin-Madison, Madison, WI 53706}

\date{\today}% It is always \today, today,

\begin{abstract}
A major challenge for ferroelectric devices is the depolarization field, which competes with and often destroys long-range polar order in the limit of ultrathin films. Recent theoretical predictions suggest a new class of materials, termed hyperferroelectics, that should be robust against the depolarization field and enable ferroelectricity down to the monolayer limit. Here we demonstrate the epitaxial growth of hexagonal LiZnSb, one of the hyperferroelectric candidate materials, by molecular-beam epitaxy on GaSb (111)B substrates. Due to the high volatility of all three atomic species, we find that LiZnSb can be grown in an adsorption-controlled window, using an excess zinc flux. Within this window, the desired polar hexagonal phase is stabilized with respect to a competing cubic polymorph, as revealed by X-ray diffraction and transmission electron microscopy measurements. First-principles calculations show that for moderate amounts of epitaxial strain and moderate concentrations of Li vacancies, the cubic LiZnSb phase is lower in formation energy than the hexagonal phase, but only by a few meV per formula unit. Therefore we suggest that kinetics plays a role in stabilizing the desired hexagonal phase at low temperatures. Our results provide a path towards experimentally demonstrating ferroelectricity and hyperferroelectricity in a new class of ternary intermetallic compounds. 
\end{abstract}

\maketitle

\section{Introduction}
To date, most studies of ferroelectric materials have concentrated on transition metal oxide systems such as PbTiO${_3}$ and BaTiO${_3}$, which feature highly ionic interactions and large Born effective charges. Unfortunately, most known ferroelectric thin films cannot maintain the desired spontaneous electric polarization with decreasing thickness \cite{junquera2003critical,sai2005ferroelectricity}. Termed ``proper ferroelectics,'' in these materials the loss of polarization results from a competing depolarization field that grows in relative strength as the material gets thinner. To overcome this challenge, recent first-principles calculations predict a new family of ferroelectric materials: $ABC$ semiconductors, also known as hexagonal Heuslers \cite{bennett2012hexagonal}. Unlike conventional proper ferroelectrics, many of these materials are predicted to be ``hyperferroelectric,'' proper ferroelectrics that can retain long-range polarization under large depolarization fields \cite{garrity2014hyperferroelectrics}. Compared with most known ferroelectrics, the hexagonal Heusler ferroelectric materials feature covalent bonding, smaller Born effective charge, and smaller band gaps \cite{garrity2014hyperferroelectrics}. Furthermore, while it is difficult to integrate oxide ferroelectrics with commonly used semiconductors \cite{demkov2014integration}, these hexagonal half-Heusler compounds are readily lattice matched to III-V semiconductors \cite{kawasaki2019heusler}. 

\begin{figure}[h!]
    \centering
    \includegraphics[width=0.45\textwidth]{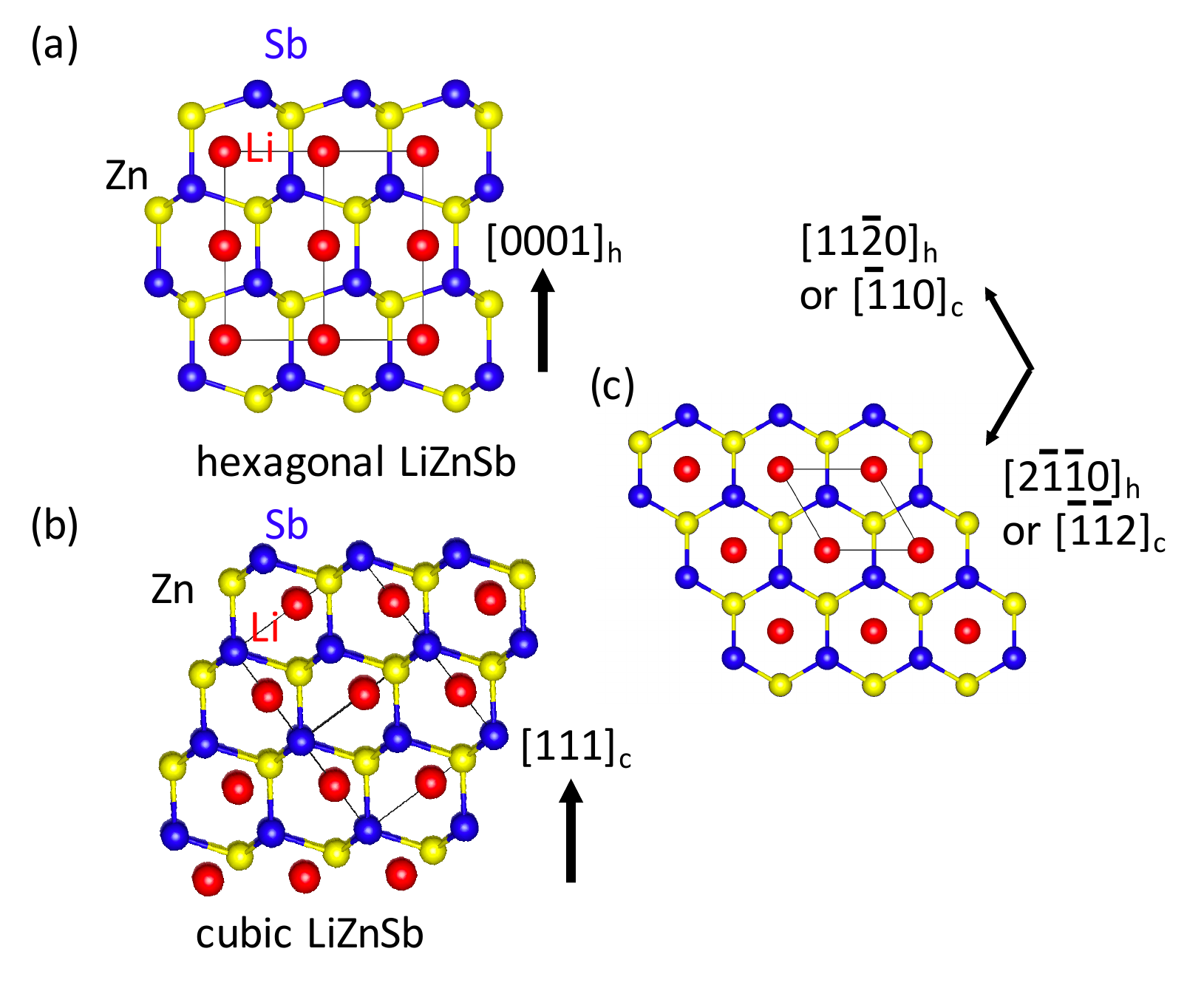}
    \caption {\textbf{Crystal structures of hexagonal and cubic LiZnSb.} (a) Hexagonal LiZnSb (\textit{LiGaGe}-type structure), which consists of a wurtzite ZnSb sublattice (yellow and blue atoms) that is stuffed with Li (red atoms). (b) Cubic LiZnSb (half Heusler structure), which consists of a zincblende ZnSb sublattice that is stuffed with Li. These two polymorphs are related by \textit{AB-AB} (hexagonal) versus \textit{ABC-ABC} (cubic) stacking along the $[0001]_{h} \parallel [111]_c$ axes. \textit{h} and \textit{c} denote cubic and hexagonal, respectively. (c) Top-down view of a single monolayer in $(0001)_h \parallel (111)_c$ orientation. In this orientation, a single monolayer of the cubic and hexagonal phase are indistinguishable. }
    \label{struc}
\end{figure}

Of the predicted compounds, hexagonal LiZnSb is one of the more promising hyperferroelectric candidate materials. Density functional theory calculations suggest LiZnSb should have a polarization of 0.56 $C/m^2$, comparable to BaTiO$_3$ \cite{garrity2014hyperferroelectrics, bennett2012hexagonal}. In this compound, the ZnSb atoms form a hexagonal wurtzite structure and the Li atoms stuff at the interstitials. A significant challenge, however, is the existence of a competing nonpolar cubic polymorph (Fig. \ref{struc}(b)), which differs in formation energy from the desired hexagonal phase by only a few meV per formula unit \cite{white2016polytype}. As such, the phase purity of LiZnSb is highly dependent on synthesis route \cite{song2019creation, white2016polytype}. Single crystalline hexagonal films, which are necessary for devices, have not yet been demonstrated. Ferroelectric switching, either in bulk or thin film form, has not yet been reported for any of the $ABC$ ferroelectric candidates.

Here, we use molecular-beam epitaxy (MBE) to demonstrate the first growth of single-crystalline, hexagonal LiZnSb thin films. Based on the high volatility of all three elements in this compound \cite{alcock1984vapour}, especially Zn, we find a low temperature ($\leq 175 \degree$C), high Zn flux regime in which the hexagonal polymorph is stabilized over competing phases. We demonstrate the epitaxial growth of LiZnSb films on GaSb (111)B with a sharp interface, as established by X-ray diffraction (XRD), reflection high-energy electron diffraction (RHEED), and scanning transmission electron microscopy (STEM). 
%\begin{figure}[H]
    %\centering
   %\includegraphics[width=0.48\textwidth]{LZS_P9.png}
    %\caption{Reflection high energy electron diffraction for beam oriented along $\langle\bar{1} 1 0\rangle_{cubic}$ / $\langle1 1 \bar{2} 0\rangle_{hex}$ (left column) and $\langle\bar{1} \bar{1} 2\rangle_{cubic}$ / $\langle 2 \bar{1} \bar{1} 0\rangle_{hex}$ (right column), respectively. (a,b) GaSb buffer layer. (c,d) LiZnSb after 30 minutes (24 nm) of growth at 190 $\degree$C with flux ratio of Li:Zn:Sb=1.07:13.00:1.00. (e,f) Epitaxial Zn cap.}
    
%\end{figure}
\section{Methods}

Epitaxial LiZnSb films were grown in a Veeco GEN 10 MBE system using the PARADIM thin film synthesis facility at Cornell University, an NSF-supported Materials Innovation Platform [www.PARADIM.org]. The typical layer structure consists of 100 nm Zn cap / 20-30 nm LiZnSb / 40-100 nm GaSb buffer / GaSb (111)B substrate, corresponding to a  3.5\% compressive lattice mismatch. Substrates were rinsed with isopropanol followed by de-ionized water and then blow dried with nitrogen before being loaded into the MBE chamber. Following thermal desorption of the native oxide under an Sb$_4$ flux, a GaSb buffer layer was grown at 490 $\degree$C, as measured by a thermocouple calibrated to the oxide-desorption temperature of GaSb. For GaSb growth we use a Sb/Ga atomic flux ratio of 3 as measured by a quartz crystal microbalance. Samples were then cooled under an Sb$_4$ flux to temperatures in the range of 100$\degree$C to 350$\degree$C, before initiating the LiZnSb growth. 

For LiZnSb growth, we use standard low-temperature effusion cells loaded with elemental Sb, elemental Zn, and a Li-Sn alloy with a starting composition of about Li$_{0.2}$Sn$_{0.8}$. The Li-Sn alloy, which consists of a mixture of Li$_2$Sn$_5$ and Sn, was used as an alternative to elemental Li due to its increased oxidation resistance. This Li-Sn alloy is prepared in a glove box, but once prepared, it can be exposed to air, greatly simplifying source loading and MBE maintenance. Since the vapor pressure of Li is more than 10$^7$ times larger than the vapor pressure of Sn at the Li-Sn cell temperature of 500$\degree$C to 670$\degree$C, we expect the Sn incorporation into our films to be negligible \cite{vaporhonig}. Due to the high relative volatility of Zn compared to Li and Sb \cite{alcock1984vapour}, we use an excess Zn/Sb atomic flux ratio of 5-25, and Li/Sb atomic flux ratios near 1. These correspond to Zn fluxes of order 10$^{14}$ to 10$^{15}$ atom/cm$^2\cdot$s, and Li and Sb fluxes of order 10$^{13}$ atom/cm$^2\cdot$s. In this regime the resulting film crystal structure is weakly dependent on relative Zn overpressure, and dependins more strongly on growth temperature and Li/Sb flux ratio. After LiZnSb growth, samples were cooled to room temperature under a Zn flux, in order to compensate for Zn desorption. Below 50$\degree$C the excess Zn begins to stick and form a cap. An epitaxial capping layer of Zn was deposited to protect the sample upon removal from vacuum.

For TEM measurements, LiZnSb cross section samples were prepared with a focused ion beam (FIB), followed by final thinning in a Fischione Model 1040 Nanomill using Ar$^+$ ions at 900V. Samples were stored in vacuum and cleaned in a GV10x DS Asher cleaner run at a power of 20 W for 10 min before being transferred into the TEM column. A probe corrected Thermo Fisher Titan STEM operated at 200 kV was used to analyze the sample. An electron probe with 24.5 mrad probe semi-convergence angle and 18.9 pA beam current was formed, achieving sub-Angstrom spatial resolution. High angle annular dark field (HAADF) images were recorded with a Fishione 3000 annular detector covering collection angle ranging from 53.9 mrad to 269.5 mrad.

We performed first-principles density functional theory (DFT) calculations in the local density approximation using ABINIT \cite{gonze2016recent}. The projector augmented wave method \cite{jollet2014generation} with pseudopotentials containing 3 valence electrons for Li ($1s^2 2s^1  2p^0$), 12 for Zn ($4s^2 3d^{10} 4p^0$), and 5 for Sb ($5s^2 5p^3$) was used. An energy cutoff of 680 eV was used for all calculations. The computed lattice constant for the cubic structure, using a $10 \times 10 \times 10$ Monkhorst-Pack $k$-point mesh, is 6.14 \AA, and the computed lattice constants for the hexagonal structure, using a $16 \times 16 \times 12$ mesh, are $a = 4.34$ \AA\ and $c = 7.03$ \AA, in good agreement with previous literatures \cite{bennett2012hexagonal, white2016polytype, toberer2009thermoelectric} and experiments \cite{white2016polytype, song2019creation, toberer2009thermoelectric, nie2014lithiation}. The effects of epitaxial strain in the (0001) plane were investigated through the strained bulk approach, with $a$ ranging from 4.08 \AA\ to 4.60 \AA, corresponding to 6\% compressive and tensile strains. We imposed the epitaxial constraint on the cubic structure by treating the cubic lattice as rhombohedral ($\alpha$ = 120$^\circ$), with a hexagonal supercell using a $16 \times 16 \times 8$ $k$-point mesh. We computed the energy to remove one Li atom from a 72 atom hexagonal supercell, with the supercells in the cubic and hexagonal structures having almost identical shapes, using a $8 \times 8 \times 4$ $k$-point mesh.

\section{Results and Discussion}

\begin{figure}[h!]
    \centering
    \includegraphics[width=0.5\textwidth]{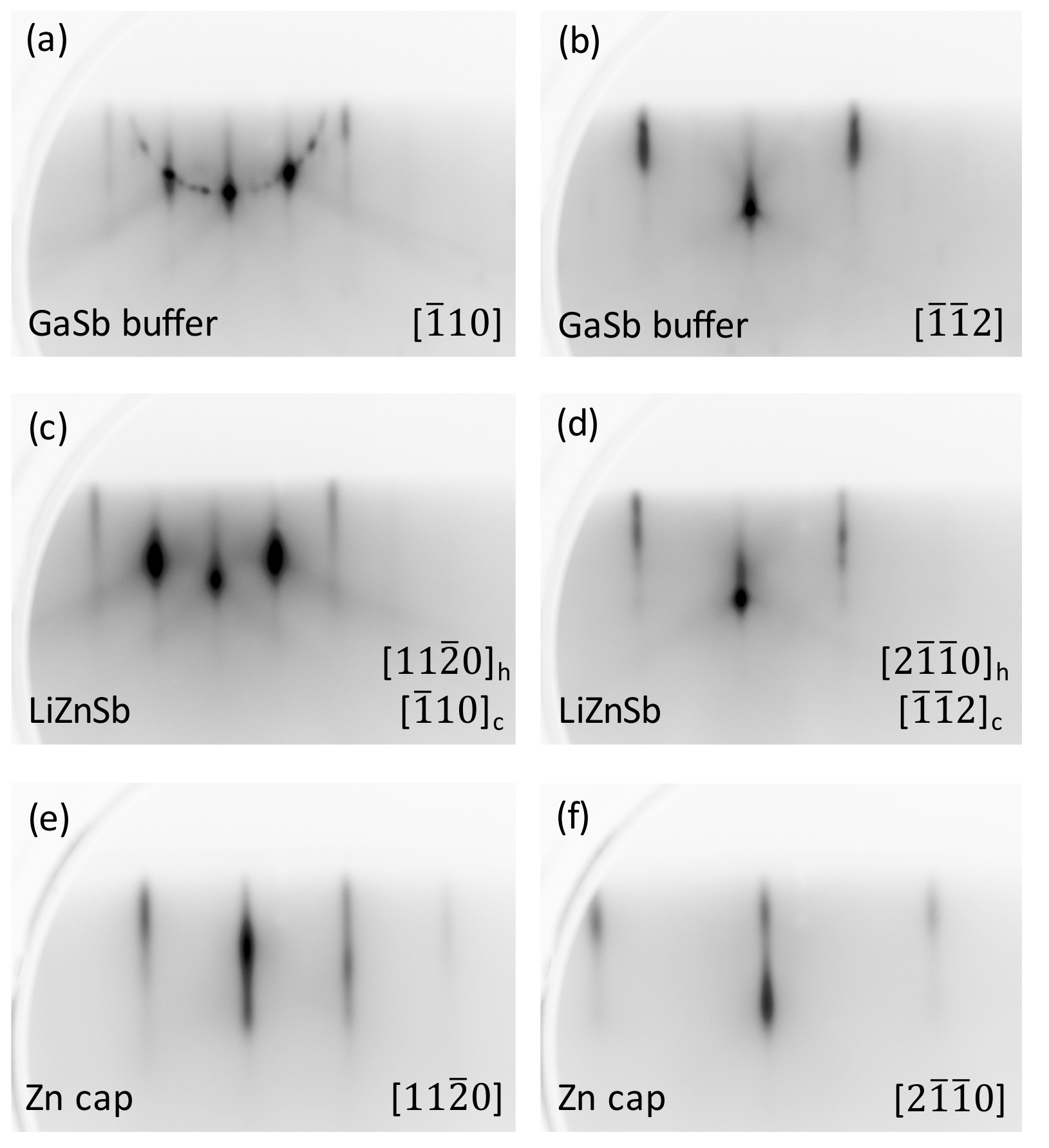}
    \caption{\textbf{Reflection high energy electron diffraction patterns at various stages for Zn-capped LiZnSb films on GaSb (111)B.} (a,b) GaSb buffer layer. (c,d) LiZnSb after 30 minutes (24 nm) of growth at 190 $\degree$C with atomic flux ratios of Li/Sb = 1.1 and Zn/Sb = 13. (e,f) Epitaxial Zn cap. Left column: electron beam oriented along $[ \bar{1} 1 0]_{c} \parallel [1 1 \bar{2} 0]_{h}$. Right column: beam oriented along $[\bar{1} \bar{1} 2]_{c} \parallel [ 2 \bar{1} \bar{1} 0]_{h}$.}
    \label{rheed}
\end{figure}

Figure \ref{rheed} shows typical RHEED patterns following the growth sequence. The GaSb buffer layer shows a sharp and streaky $(1 \times 12)$ pattern, indicative of smooth growth (Figs. \ref{rheed}(a) and \ref{rheed}(b)). For the LiZnSb layers (Figs. \ref{rheed}(c) and \ref{rheed}(d)), sharp and streaky $(1\times 1)$ patterns are observed over a wide range of Li/Sb flux ratios (0.4 to 2) and growth temperatures (125 to 350 $\degree$C). For growth temperatures above 225 $\degree$C, even though the RHEED shows a sharp and streaky $(1\times 1)$ pattern indicative of changes in surface termination, no bulk reflections from Li-Zn-Sb phases are observed by post-growth X-ray diffraction, indicating minimal LiZnSb sticking on the surface at elevated temperatures. By lowering the temperature below 225 $\degree$C, XRD signals from film reflections appear.

\begin{figure}[h!]
    \centering
    \includegraphics[width=0.5\textwidth]{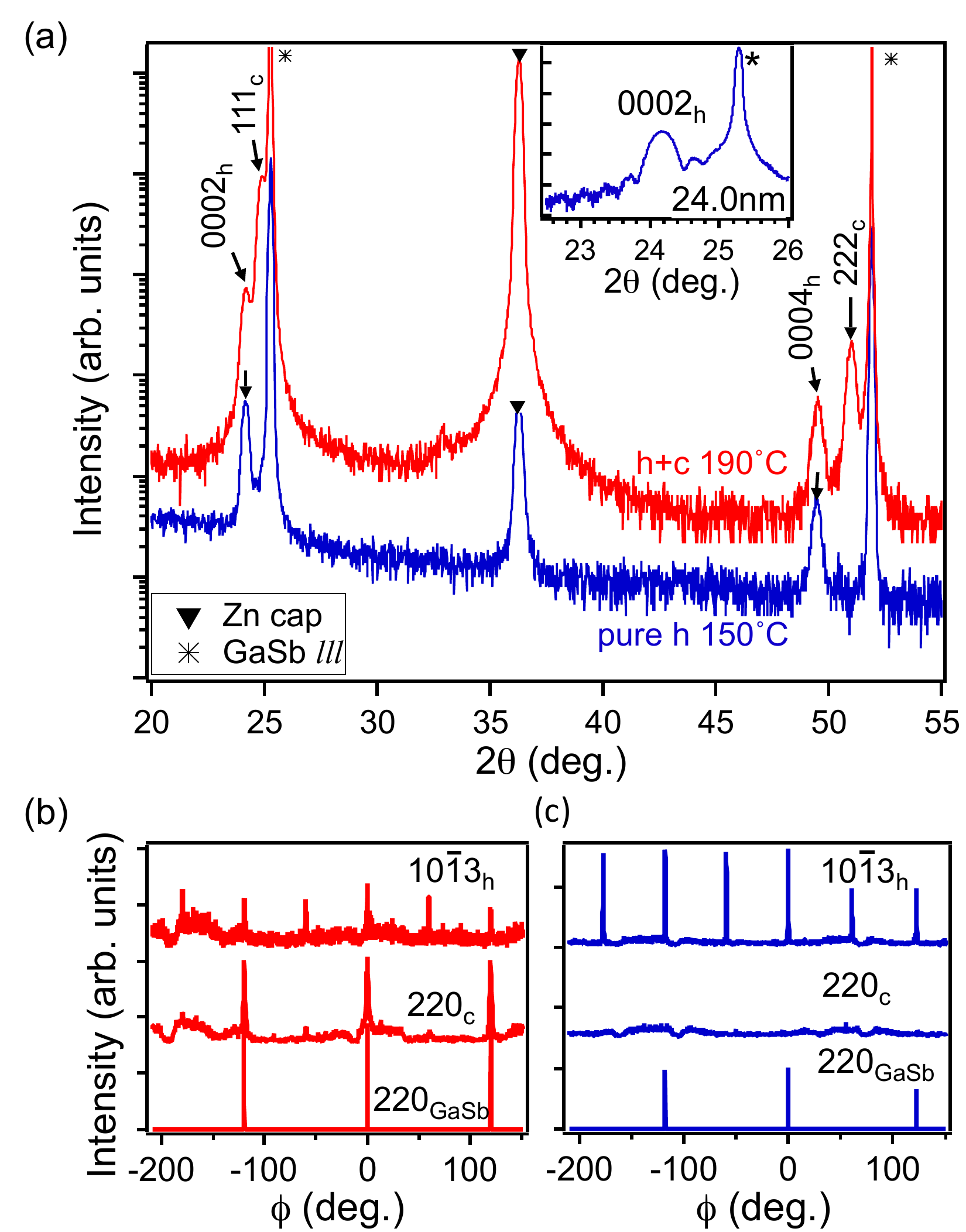}
    \caption{\textbf{Distinguishing hexagonal from cubic LiZnSb by x-ray diffraction (Cu K$_{\alpha}$).} (a) Out-of-plane $\theta-2\theta$ scans for samples grown at two different temperatures. The sample grown at T$_{growth}$=190 $\degree$C (red curve) shows a mixture of hexagonal $000l_h$ and cubic $lll_c$ reflections. The sample growth at a lower temperature of 150 $\degree$C (blue curve), shows only hexagonal $000l_h$ reflections. The inset shows a zoom-in near the $0002_{h}$ reflection of the T$_{growth}$=150 $\degree$C sample. The Kiessig fringe spacing corresponds to a thickness of 24 nm. (b) In-plane $\phi$ scans of the T$_{growth}$=190 $\degree$C mixed-phase sample. Both hexagonal $10\bar{1}3_h$ and cubic $220_c$ LiZnSb reflections for are present. (c) In-plane $\phi$ scans of the pure hexagonal LiZnSb sample grown at T$_{growth}$=150 $\degree$C.}
    \label{xrd}
\end{figure}

\begin{figure*}
    \centering
    \includegraphics[width=1\textwidth]{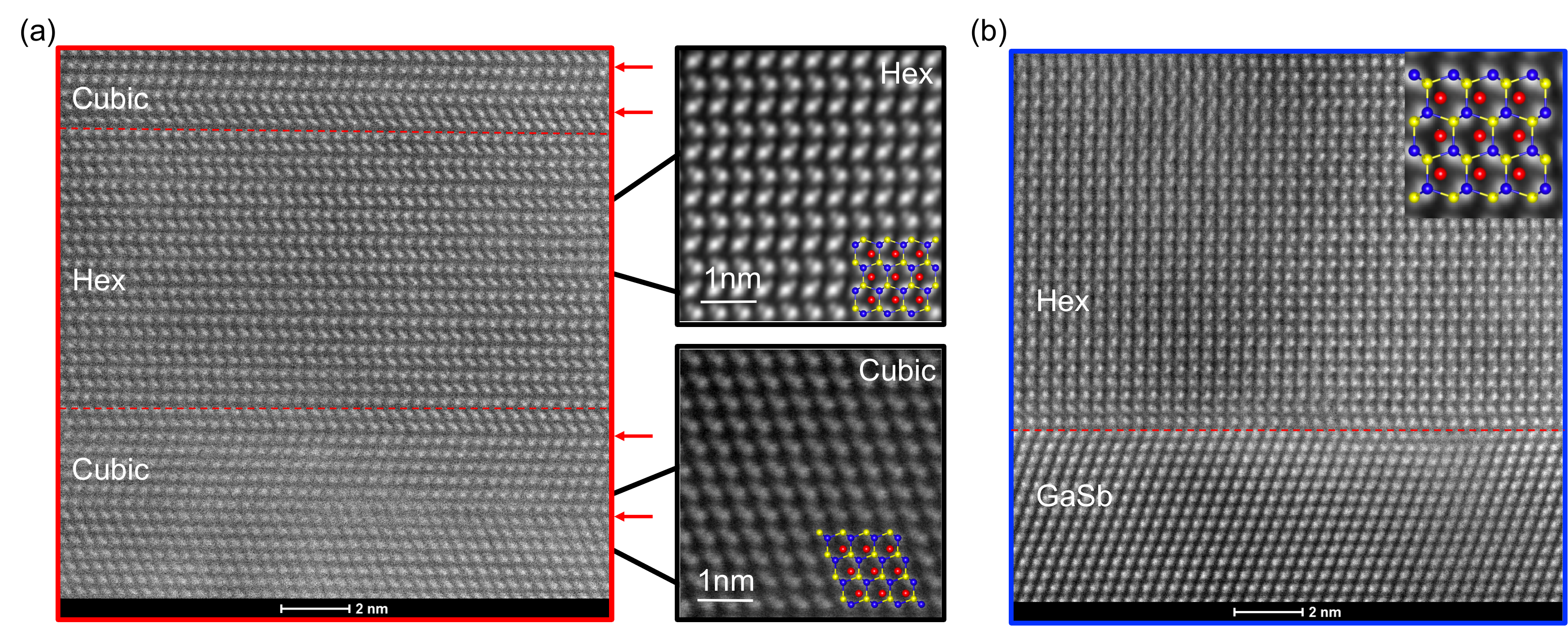}
    \caption{\textbf{Cross-sectional STEM along a hexagonal $[11\bar{2}0]$ zone axis. The growth direction points upwards.} (a) Mixed-phase sample. In the STEM image on the left side, distinct layers with the hexagonal structure and the cubic structure are observed. Red lines denote the interfaces between the hexagonal and cubic regions. Arrows denote stacking faults. On the right side, magnified STEM images of the cubic structure and the hexagonal structure are displayed. (b) STEM image of a phase-pure hexagonal LiZnSb sample at the interface between the GaSb substrate and hexagonal LiZnSb. The inset is a high resolution STEM image of the hexagonal phase. Due to the light mass of Li atoms, they cannot be detected by STEM, so we can only see Zn and Sb atoms. The schematic crystal structures are placed on top of the STEM images and the color coding is as follows:  red spheres are Li, yellow spheres are Zn, and blue spheres are Sb.}
    \label{stem}
\end{figure*} 

Figure \ref{xrd}(a) shows the XRD patterns (Cu K$_\alpha$) for two samples, one grown at $190\degree$C and the other grown at $150 \degree$C. For the higher temperature sample, two sets of reflections are observed, corresponding to cubic $lll_c$ reflections and hexagonal $000l_h$ reflections. For the sample grown at lower temperature, only one set of reflections is observed. In the $190 \degree$C sample, the lower angle reflections ($2\theta = 24.29 \degree$ and $49.63 \degree$) correspond to an out-of-plane $d_\perp$ spacing of 7.34 \AA, and the higher angle reflections ($2\theta = 24.93 \degree$ and 51.06$ \degree$) correspond to $d_\perp=7.13$ \AA. In comparison, previous measurements of bulk cubic LiZnSb report $2d_{111}=a \frac{2\sqrt 3}{3}=7.19$ \AA\ ($a=6.23$ \AA\ \cite{white2016polytype,white2018expanding}). For bulk hexagonal LiZnSb the experimental lattice parameters range from $c=7.15$ \AA\ to 7.24 \AA\ depending on Li stoichiometry \cite{nie2014lithiation, song2019creation, toberer2009thermoelectric}, and  $c=6.02$ \AA\ for 2D-ZnSb \cite{song2019creation}. Our measured values of $d_\perp$ fall within the range of these reports, and therefore we cannot make an assignment of cubic versus hexagonal reflections from the magnitudes of $d_\perp$ alone.

To distinguish cubic from hexagonal LiZnSb, we perform in-plane $\phi$ scans (Figs. \ref{xrd}(b) and \ref{xrd}(c)). For the higher growth temperature sample we observe both the cubic $220_c$ and hexagonal $10\bar{1}3_h$ LiZnSb reflections, while for the lower temperature sample we observe only the hexagonal $10\bar{1}3_h$. From these measurements, we determine that the lower angle reflections correspond to the hexagonal phase with $c=7.34$ \AA, while the higher angle reflections correspond to the cubic phase with $2 d_{\perp,111}=7.13$ \AA. Projecting the measured $10\bar{1}3_h$ and $220_c$ reflections to the growth plane, we find in-plane lattice parameters of $a=4.43$ \AA\ for hexagonal and $d_{\parallel,110}=4.39$ \AA\ for cubic, respectively. These measurements are in good agreement with previous measurements on bulk samples, which report $a=4.43$ \AA\ for hexagonal \cite{toberer2009thermoelectric, white2016polytype, song2019creation} and $d_{110}=a\sqrt 2=4.41$ \AA\ for cubic \cite{white2016polytype, white2018expanding}. For the single-phase hexagonal film grown at $150 \degree$C, we also observe finite thickness fringes in the $2\theta$ scan (Fig. \ref{xrd}(a), insert), indicative of sharp interfaces between film and substrate. These results suggest that lowering the growth temperature produces phase-pure hexagonal LiZnSb films, while higher temperature growth results in a mixture of cubic and hexagonal polymorphs. 

Our assignment of cubic and hexagonal LiZnSb is corroborated by cross-sectional scanning transmission electron microscopy (STEM). For the higher temperature sample (Fig. \ref{stem}(a)) we observe regions of both $ABC-ABC$ and $AB-AB$ stacking along the growth direction, corresponding to cubic and hexagonal phases, respectively. In contrast, for the low temperature sample we observe only the hexagonal phase with $AB-AB$ stacking (Fig. \ref{stem}(b)). STEM of this phase-pure hexagonal sample also shows a sharp interface between the hexagonal LiZnSb film and the cubic GaSb (111)B substrate, consistent with the sharp Kiessig fringes observed by XRD.

\begin{figure}
    \centering
    \includegraphics[width=0.5\textwidth]{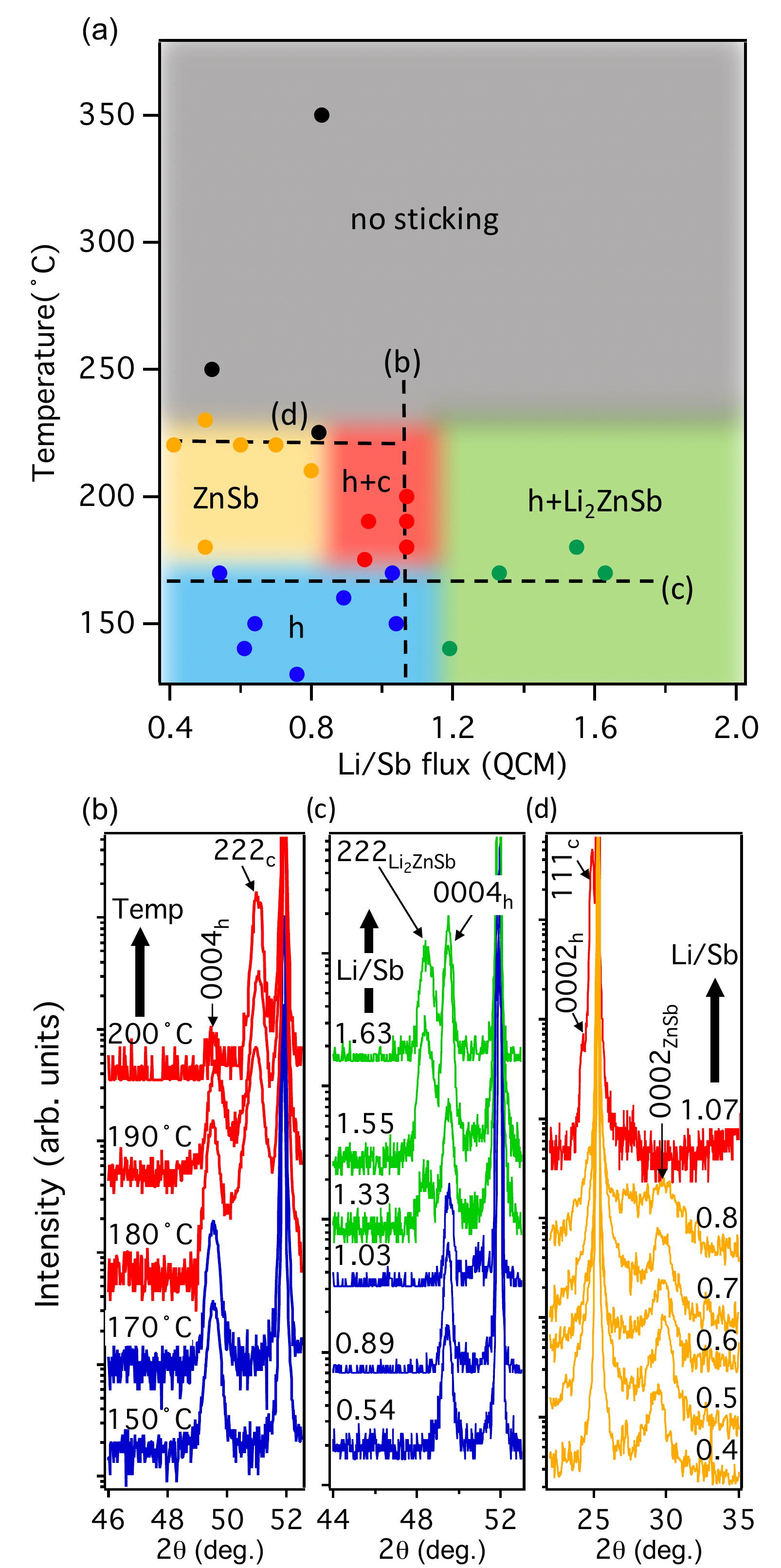}
    \caption{(a) Diagram showing what phases form in thin films grown on GaSb (111)B at different growth temperatures and Li/Sb flux ratios. \textit{``c''} and \textit{``h''} represent the cubic and hexagonal LiZnSb, respectively. (b) $\theta-2\theta$ scans around the GaSb 222 reflection of 20-30 nm thick films grown at fixed Li/Sb flux ratio in the range of 1.03 to 1.07, with varying growth temperature. The corresponding cut through the growth parameter diagram is denoted by a dashed line in (a). (c) $\theta-2\theta$ scans near the GaSb 222 reflection for samples grown at fixed temperature ($167 \pm 7\degree$C), with varying Li/Sb flux ratios. (d) $\theta-2\theta$ scans near the GaSb 222 reflection for ZnSb samples grown at fixed temperature ($212 \pm 12\degree$C), for varying Li/Sb flux ratios.}
    \label{phasediagram}
\end{figure}

The phase diagram for MBE growth is summarized in figure 5(a). For a finite region centered near a Li/Sb flux ratio of 1, we find that decreasing the substrate temperature below 175 $\degree$C favors the formation of pure-phase hexagonal films (Fig. 5(b)). For fixed growth temperatures below 175 $\degree$C, increasing the Li/Sb flux ratio beyond 1 leads to the formation of Li$_2$ZnSb with the cubic full Heusler structure (Fig. 5(c)). For growth at moderate temperatures of 200-225$\degree$C, decreasing the Li/Sb flux ratio 0.8 lead to the formation of hexagonal ZnSb (Fig. 5(d)), the same phase found by Li de-intercalation of hexagonal LiZnSb \cite{song2019creation}.

This result is somewhat surprising in light of our DFT calculations for bulk LiZnSb, which show that the cubic phase has lower energy than the hexagonal phase by about 35 meV per formula unit, comparable to the previously reported value of 30 meV \cite{white2016polytype}, and hence the cubic is the expected stable phase. The results are similar using both LDA (local density approximation) and GGA (generalized gradient approximation) functionals \cite{white2018expanding}. Note, however, that 30-35 meV per formula unit is similar in magnitude to the thermal energy, $k_B T = 40$meV, at a growth temperature of $200\degree$C. Therefore we do not expect a strong thermodynamic driving force to prefer one phase over the other. Although a mixture of hexagonal and cubic polymorphs is seen at higher growth temperatures, at low temperature only the hexagonal polymorph is seen.

The formation of the higher-energy hexagonal phase does not appear to be related to epitaxial strain stabilization. In-plane lattice parameters of our MBE grown films appear to be relaxed from that of the GaSb substrate (a = 4.43 \AA\ for hexagonal LiZnSb, $d_{\parallel, 110} = 4.39$ \AA\ for cubic LiZnSb, and $d_{\parallel, 110} = 4.31$ \AA\ for the GaSb substrate). Furthermore, our DFT calculations for the cubic-hexagonal energy difference for epitaxially strained films as a function of strain show that the energy difference of 35 meV/fu is quite insensitive to strain in the range from -6\% to +6\%.

Finally, we checked the effects of Li stoichiometry on the cubic-hexagonal energy difference. Previous Li de-intercalation studies of LiZnSb suggest that LiZnSb is stable over a range of Li composition \cite{song2019creation}, ranging from stoichiometric LiZnSb to the layered 2D polymorph of ZnSb. From first-principles calculations for the relative energy of the cubic and hexagonal phases with one Li removed from each 72 atom supercell, we find that while it is slightly more energetically favorable to form a Li vacancy in the hexagonal structure than in the cubic structure, a vacancy concentration of approximately 50\% would be required for this energy to stabilize the hexagonal phase relative to the cubic phase. Our experimental lattice parameter of $c = 7.34$ \AA\ is much closer to that of nominally stoichiometric bulk LiZnSb ($c = 7.24$ \AA) than to that of 2D ZnSb ($c = 6.02$ \AA). Therefore it is unlikely that Li vacancies are responsible for stabilizing the hexagonal phase.

Given the very small difference in formation energies for cubic and hexagonal compared to $k_B T$, and relative insensitivity to strain and Li stoichiometry, the most likely reason for stabilizing the hexagonal phase at low temperature is kinetics. In support of this idea, we find that after extended exposure to the 200 keV electron beam during TEM measurements, the relative volume fraction of cubic to hexagonal phase increases. Recent wet synthesis of LiZnSb also suggests that kinetics plays a strong role, as the hexagonal phase is favored at lower temperatures and shorter times, while the cubic phase is favored at higher temperatures and longer times \cite{white2018expanding}. Note, however, that the kinetic pathway for a wet synthesis is very different than epitaxy from vapor during MBE growth. Although they are often small, finite temperature effects or changes in the Zn chemical potential may modify the true formation energies sufficiently to change phase selection in this case.

\section{Conclusion}
In this paper, we present the first epitaxial growth of LiZnSb thin films and showed a wide adsorption-controlled growth window in which the hexagonal phase is stabilized with MBE. This study of the MBE growth of LiZnSb provides solutions to the obstacles in growing single crystalline epitaxial films of $ABC$ hyperferroelectric candidates \cite{bennett2012hexagonal,garrity2014hyperferroelectrics} which are composed entirely of elements with relatively high vapor pressures. When combined with metallic $ABC$ films, e.g., LaPtSb and LaAuSb \cite{du2019high, strohbeen2019electronically}, the family of hexagonal Heuslers provides a platform for all-epitaxial ferroelectric and polar metal heterostructures.

\section{Acknowledgments}

This work was supported by the United States Army Research Office (ARO Award number W911NF-17-1-0254). Synthesis efforts at the PARADIM facility were supported by the National Science Foundation (Platform for the Accelerated Realization, Analysis, and Discovery of Interface Materials (PARADIM)) under Cooperative Agreement No. DMR-1539918. High-resolution STEM characterization was supported by the Department of Energy Basic Energy Science (DE-FG02-08ER46547), with facilities supported by Wisconsin MRSEC (DMR-1720415). DFT calculations were supported by the Office of Naval Research (ONR N00014-17-1-2770).

\bibliographystyle{apsrev}
\bibliography{bibliography}

\begin{thebibliography}{17}
\expandafter\ifx\csname natexlab\endcsname\relax\def\natexlab#1{#1}\fi
\expandafter\ifx\csname bibnamefont\endcsname\relax
  \def\bibnamefont#1{#1}\fi
\expandafter\ifx\csname bibfnamefont\endcsname\relax
  \def\bibfnamefont#1{#1}\fi
\expandafter\ifx\csname citenamefont\endcsname\relax
  \def\citenamefont#1{#1}\fi
\expandafter\ifx\csname url\endcsname\relax
  \def\url#1{\texttt{#1}}\fi
\expandafter\ifx\csname urlprefix\endcsname\relax\def\urlprefix{URL }\fi
\providecommand{\bibinfo}[2]{#2}
\providecommand{\eprint}[2][]{\url{#2}}

\bibitem[{\citenamefont{Junquera and Ghosez}(2003)}]{junquera2003critical}
\bibinfo{author}{\bibfnamefont{J.}~\bibnamefont{Junquera}} \bibnamefont{and}
  \bibinfo{author}{\bibfnamefont{P.}~\bibnamefont{Ghosez}},
  \bibinfo{journal}{Nature} \textbf{\bibinfo{volume}{422}},
  \bibinfo{pages}{506} (\bibinfo{year}{2003}).

\bibitem[{\citenamefont{Sai et~al.}(2005)\citenamefont{Sai, Kolpak, and
  Rappe}}]{sai2005ferroelectricity}
\bibinfo{author}{\bibfnamefont{N.}~\bibnamefont{Sai}},
  \bibinfo{author}{\bibfnamefont{A.~M.} \bibnamefont{Kolpak}},
  \bibnamefont{and} \bibinfo{author}{\bibfnamefont{A.~M.} \bibnamefont{Rappe}},
  \bibinfo{journal}{Physical Review B} \textbf{\bibinfo{volume}{72}},
  \bibinfo{pages}{020101} (\bibinfo{year}{2005}).

\bibitem[{\citenamefont{Bennett et~al.}(2012)\citenamefont{Bennett, Garrity,
  Rabe, and Vanderbilt}}]{bennett2012hexagonal}
\bibinfo{author}{\bibfnamefont{J.~W.} \bibnamefont{Bennett}},
  \bibinfo{author}{\bibfnamefont{K.~F.} \bibnamefont{Garrity}},
  \bibinfo{author}{\bibfnamefont{K.~M.} \bibnamefont{Rabe}}, \bibnamefont{and}
  \bibinfo{author}{\bibfnamefont{D.}~\bibnamefont{Vanderbilt}},
  \bibinfo{journal}{Physical review letters} \textbf{\bibinfo{volume}{109}},
  \bibinfo{pages}{167602} (\bibinfo{year}{2012}).

\bibitem[{\citenamefont{Garrity et~al.}(2014)\citenamefont{Garrity, Rabe, and
  Vanderbilt}}]{garrity2014hyperferroelectrics}
\bibinfo{author}{\bibfnamefont{K.~F.} \bibnamefont{Garrity}},
  \bibinfo{author}{\bibfnamefont{K.~M.} \bibnamefont{Rabe}}, \bibnamefont{and}
  \bibinfo{author}{\bibfnamefont{D.}~\bibnamefont{Vanderbilt}},
  \bibinfo{journal}{Physical review letters} \textbf{\bibinfo{volume}{112}},
  \bibinfo{pages}{127601} (\bibinfo{year}{2014}).

\bibitem[{\citenamefont{Demkov and Posadas}(2014)}]{demkov2014integration}
\bibinfo{author}{\bibfnamefont{A.~A.} \bibnamefont{Demkov}} \bibnamefont{and}
  \bibinfo{author}{\bibfnamefont{A.~B.} \bibnamefont{Posadas}},
  \emph{\bibinfo{title}{Integration of functional oxides with semiconductors}},
  vol.~\bibinfo{volume}{25} (\bibinfo{publisher}{Springer},
  \bibinfo{year}{2014}).

\bibitem[{\citenamefont{Kawasaki}(2019)}]{kawasaki2019heusler}
\bibinfo{author}{\bibfnamefont{J.~K.} \bibnamefont{Kawasaki}},
  \bibinfo{journal}{APL Materials} \textbf{\bibinfo{volume}{7}},
  \bibinfo{pages}{080907} (\bibinfo{year}{2019}).

\bibitem[{\citenamefont{White et~al.}(2016)\citenamefont{White, Miller, and
  Vela}}]{white2016polytype}
\bibinfo{author}{\bibfnamefont{M.~A.} \bibnamefont{White}},
  \bibinfo{author}{\bibfnamefont{G.~J.} \bibnamefont{Miller}},
  \bibnamefont{and} \bibinfo{author}{\bibfnamefont{J.}~\bibnamefont{Vela}},
  \bibinfo{journal}{J. Am. Chem. Soc.} \textbf{\bibinfo{volume}{138}},
  \bibinfo{pages}{14574} (\bibinfo{year}{2016}).

\bibitem[{\citenamefont{Song et~al.}(2019)\citenamefont{Song, Song, Wang, Lee,
  Hwang, Lee, Lee, Kim, Lee, Kim et~al.}}]{song2019creation}
\bibinfo{author}{\bibfnamefont{J.}~\bibnamefont{Song}},
  \bibinfo{author}{\bibfnamefont{H.~Y.} \bibnamefont{Song}},
  \bibinfo{author}{\bibfnamefont{Z.}~\bibnamefont{Wang}},
  \bibinfo{author}{\bibfnamefont{S.}~\bibnamefont{Lee}},
  \bibinfo{author}{\bibfnamefont{J.-Y.} \bibnamefont{Hwang}},
  \bibinfo{author}{\bibfnamefont{S.~Y.} \bibnamefont{Lee}},
  \bibinfo{author}{\bibfnamefont{J.}~\bibnamefont{Lee}},
  \bibinfo{author}{\bibfnamefont{D.}~\bibnamefont{Kim}},
  \bibinfo{author}{\bibfnamefont{K.~H.} \bibnamefont{Lee}},
  \bibinfo{author}{\bibfnamefont{Y.}~\bibnamefont{Kim}}, \bibnamefont{et~al.},
  \bibinfo{journal}{Science advances} \textbf{\bibinfo{volume}{5}},
  \bibinfo{pages}{eaax0390} (\bibinfo{year}{2019}).

\bibitem[{\citenamefont{Alcock et~al.}(1984)\citenamefont{Alcock, Itkin, and
  Horrigan}}]{alcock1984vapour}
\bibinfo{author}{\bibfnamefont{C.}~\bibnamefont{Alcock}},
  \bibinfo{author}{\bibfnamefont{V.}~\bibnamefont{Itkin}}, \bibnamefont{and}
  \bibinfo{author}{\bibfnamefont{M.}~\bibnamefont{Horrigan}},
  \bibinfo{journal}{Canadian Metallurgical Quarterly}
  \textbf{\bibinfo{volume}{23}}, \bibinfo{pages}{309} (\bibinfo{year}{1984}).

\bibitem[{\citenamefont{Honig and Kramer}(1969)}]{vaporhonig}
\bibinfo{author}{\bibfnamefont{R.~E.} \bibnamefont{Honig}} \bibnamefont{and}
  \bibinfo{author}{\bibfnamefont{D.~A.} \bibnamefont{Kramer}},
  \bibinfo{journal}{RCA Review} \textbf{\bibinfo{volume}{30}},
  \bibinfo{pages}{285} (\bibinfo{year}{1969}).

\bibitem[{\citenamefont{Gonze et~al.}(2016)\citenamefont{Gonze, Jollet, Araujo,
  Adams, Amadon, Applencourt, Audouze, Beuken, Bieder, Bokhanchuk
  et~al.}}]{gonze2016recent}
\bibinfo{author}{\bibfnamefont{X.}~\bibnamefont{Gonze}},
  \bibinfo{author}{\bibfnamefont{F.}~\bibnamefont{Jollet}},
  \bibinfo{author}{\bibfnamefont{F.~A.} \bibnamefont{Araujo}},
  \bibinfo{author}{\bibfnamefont{D.}~\bibnamefont{Adams}},
  \bibinfo{author}{\bibfnamefont{B.}~\bibnamefont{Amadon}},
  \bibinfo{author}{\bibfnamefont{T.}~\bibnamefont{Applencourt}},
  \bibinfo{author}{\bibfnamefont{C.}~\bibnamefont{Audouze}},
  \bibinfo{author}{\bibfnamefont{J.-M.} \bibnamefont{Beuken}},
  \bibinfo{author}{\bibfnamefont{J.}~\bibnamefont{Bieder}},
  \bibinfo{author}{\bibfnamefont{A.}~\bibnamefont{Bokhanchuk}},
  \bibnamefont{et~al.}, \bibinfo{journal}{Computer Physics Communications}
  \textbf{\bibinfo{volume}{205}}, \bibinfo{pages}{106} (\bibinfo{year}{2016}).

\bibitem[{\citenamefont{Jollet et~al.}(2014)\citenamefont{Jollet, Torrent, and
  Holzwarth}}]{jollet2014generation}
\bibinfo{author}{\bibfnamefont{F.}~\bibnamefont{Jollet}},
  \bibinfo{author}{\bibfnamefont{M.}~\bibnamefont{Torrent}}, \bibnamefont{and}
  \bibinfo{author}{\bibfnamefont{N.}~\bibnamefont{Holzwarth}},
  \bibinfo{journal}{Computer Physics Communications}
  \textbf{\bibinfo{volume}{185}}, \bibinfo{pages}{1246} (\bibinfo{year}{2014}).

\bibitem[{\citenamefont{Toberer et~al.}(2009)\citenamefont{Toberer, May,
  Scanlon, and Snyder}}]{toberer2009thermoelectric}
\bibinfo{author}{\bibfnamefont{E.~S.} \bibnamefont{Toberer}},
  \bibinfo{author}{\bibfnamefont{A.~F.} \bibnamefont{May}},
  \bibinfo{author}{\bibfnamefont{C.~J.} \bibnamefont{Scanlon}},
  \bibnamefont{and} \bibinfo{author}{\bibfnamefont{G.~J.}
  \bibnamefont{Snyder}}, \bibinfo{journal}{Journal of Applied Physics}
  \textbf{\bibinfo{volume}{105}}, \bibinfo{pages}{063701}
  (\bibinfo{year}{2009}).

\bibitem[{\citenamefont{Nie et~al.}(2014)\citenamefont{Nie, Cheng, Zhu,
  Asayesh-Ardakani, Tao, Mashayek, Han, Schwingenschlogl, Klie, Vaddiraju
  et~al.}}]{nie2014lithiation}
\bibinfo{author}{\bibfnamefont{A.}~\bibnamefont{Nie}},
  \bibinfo{author}{\bibfnamefont{Y.}~\bibnamefont{Cheng}},
  \bibinfo{author}{\bibfnamefont{Y.}~\bibnamefont{Zhu}},
  \bibinfo{author}{\bibfnamefont{H.}~\bibnamefont{Asayesh-Ardakani}},
  \bibinfo{author}{\bibfnamefont{R.}~\bibnamefont{Tao}},
  \bibinfo{author}{\bibfnamefont{F.}~\bibnamefont{Mashayek}},
  \bibinfo{author}{\bibfnamefont{Y.}~\bibnamefont{Han}},
  \bibinfo{author}{\bibfnamefont{U.}~\bibnamefont{Schwingenschlogl}},
  \bibinfo{author}{\bibfnamefont{R.~F.} \bibnamefont{Klie}},
  \bibinfo{author}{\bibfnamefont{S.}~\bibnamefont{Vaddiraju}},
  \bibnamefont{et~al.}, \bibinfo{journal}{Nano letters}
  \textbf{\bibinfo{volume}{14}}, \bibinfo{pages}{5301} (\bibinfo{year}{2014}).

\bibitem[{\citenamefont{White et~al.}(2018)\citenamefont{White, Baumler, Chen,
  Venkatesh, Medina-Gonzalez, Rossini, Zaikina, Chan, and
  Vela}}]{white2018expanding}
\bibinfo{author}{\bibfnamefont{M.~A.} \bibnamefont{White}},
  \bibinfo{author}{\bibfnamefont{K.~J.} \bibnamefont{Baumler}},
  \bibinfo{author}{\bibfnamefont{Y.}~\bibnamefont{Chen}},
  \bibinfo{author}{\bibfnamefont{A.}~\bibnamefont{Venkatesh}},
  \bibinfo{author}{\bibfnamefont{A.~M.} \bibnamefont{Medina-Gonzalez}},
  \bibinfo{author}{\bibfnamefont{A.~J.} \bibnamefont{Rossini}},
  \bibinfo{author}{\bibfnamefont{J.~V.} \bibnamefont{Zaikina}},
  \bibinfo{author}{\bibfnamefont{E.~M.} \bibnamefont{Chan}}, \bibnamefont{and}
  \bibinfo{author}{\bibfnamefont{J.}~\bibnamefont{Vela}},
  \bibinfo{journal}{Chemistry of Materials} \textbf{\bibinfo{volume}{30}},
  \bibinfo{pages}{6173} (\bibinfo{year}{2018}).

\bibitem[{\citenamefont{Du et~al.}(2019)\citenamefont{Du, Lim, Zhang,
  Strohbeen, Shourov, Rodolakis, McChesney, Voyles, Fredrickson, and
  Kawasaki}}]{du2019high}
\bibinfo{author}{\bibfnamefont{D.}~\bibnamefont{Du}},
  \bibinfo{author}{\bibfnamefont{A.}~\bibnamefont{Lim}},
  \bibinfo{author}{\bibfnamefont{C.}~\bibnamefont{Zhang}},
  \bibinfo{author}{\bibfnamefont{P.~J.} \bibnamefont{Strohbeen}},
  \bibinfo{author}{\bibfnamefont{E.~H.} \bibnamefont{Shourov}},
  \bibinfo{author}{\bibfnamefont{F.}~\bibnamefont{Rodolakis}},
  \bibinfo{author}{\bibfnamefont{J.~L.} \bibnamefont{McChesney}},
  \bibinfo{author}{\bibfnamefont{P.}~\bibnamefont{Voyles}},
  \bibinfo{author}{\bibfnamefont{D.~C.} \bibnamefont{Fredrickson}},
  \bibnamefont{and} \bibinfo{author}{\bibfnamefont{J.~K.}
  \bibnamefont{Kawasaki}}, \bibinfo{journal}{APL Materials}
  \textbf{\bibinfo{volume}{7}}, \bibinfo{pages}{121107} (\bibinfo{year}{2019}).

\bibitem[{\citenamefont{Strohbeen et~al.}(2019)\citenamefont{Strohbeen, Du,
  Zhang, Shourov, Rodolakis, McChesney, Voyles, and
  Kawasaki}}]{strohbeen2019electronically}
\bibinfo{author}{\bibfnamefont{P.~J.} \bibnamefont{Strohbeen}},
  \bibinfo{author}{\bibfnamefont{D.}~\bibnamefont{Du}},
  \bibinfo{author}{\bibfnamefont{C.}~\bibnamefont{Zhang}},
  \bibinfo{author}{\bibfnamefont{E.~H.} \bibnamefont{Shourov}},
  \bibinfo{author}{\bibfnamefont{F.}~\bibnamefont{Rodolakis}},
  \bibinfo{author}{\bibfnamefont{J.~L.} \bibnamefont{McChesney}},
  \bibinfo{author}{\bibfnamefont{P.~M.} \bibnamefont{Voyles}},
  \bibnamefont{and} \bibinfo{author}{\bibfnamefont{J.~K.}
  \bibnamefont{Kawasaki}}, \bibinfo{journal}{Physical Review Materials}
  \textbf{\bibinfo{volume}{3}}, \bibinfo{pages}{024201} (\bibinfo{year}{2019}).

\end{thebibliography}
\end{document}